\documentclass[%
rsi, reprint%
 ,twocolumn%
 ,secnumarabic%
,amssymb, amsmath, nobibnotes, floatfix, citeautoscript,aip]{revtex4-1}
\usepackage{bm}%
\usepackage[T1]{fontenc}
\expandafter\ifx\csname package@font\endcsname\relax\else
 \expandafter\expandafter
 \expandafter\usepackage
 \expandafter\expandafter
 \expandafter{\csname package@font\endcsname}%
\fi
\ifx\pdfoutput\undefined
\usepackage{graphicx}
\else
\usepackage[pdftex]{graphicx}
\usepackage{epstopdf}
\fi
\usepackage{hyperref}

\hypersetup{
    bookmarks=false,         
    unicode=false,          
    pdftoolbar=false,        
    pdfmenubar=false,        
    pdffitwindow=false,     
    pdfstartview={FitH},    
    pdfnewwindow=true,      
    colorlinks=true,       
    linkcolor=black,          
    citecolor=blue,        
    filecolor=blue,      
    urlcolor=blue           
}

\usepackage{natbib}
\bibpunct{(}{)}{,}{s}{,}{,}

\begin{document}
\bibliographystyle{ieeetr}
\makeatletter 
\renewcommand\@biblabel[1]{$^{#1}$} 
\makeatother

\title{A near-field scanning microwave microscope based on a superconducting resonator for low power measurements}
\author{S. E. de Graaf}
\email{degraaf@chalmers.se}
\author{A. V. Danilov}
\author{A. Adamyan}
\author{S. E. Kubatkin}

\affiliation{Department of Microtechnology and Nanoscience, MC2, Chalmers University of Technology, SE-41296 Goteborg, Sweden}
\date{\today}

\begin{abstract}
We report on the design and performance of a cryogenic (300 mK) near-field scanning microwave microscope. It uses a microwave resonator as the near-field sensor, operating at a frequency of 6 GHz and microwave probing amplitudes down to 100 $\rm{\mu V}$, approaching low enough photon population ($N\sim 1000$) of the resonator such that coherent quantum manipulation becomes feasible. The resonator is made out of a miniaturized distributed fractal superconducting circuit that is integrated with the probing tip, micromachined to be compact enough such that it can be mounted directly on a quartz tuning-fork, and used for parallel operation as an atomic force microscope (AFM). The resonator is magnetically coupled to a transmission line for readout, and to achieve enhanced sensitivity we employ a Pound-Drever-Hall measurement scheme to lock to the resonance frequency. We achieve a well localized near-field around the tip such that the microwave resolution is comparable to the AFM resolution, and a capacitive sensitivity down to $6.4\cdot 10^{-20}$ F$/\sqrt{\rm{Hz}}$, limited by mechanical noise. We believe that the results presented here are a significant step towards probing quantum systems at the nanoscale using near-field scanning microwave microscopy.
\end{abstract}
\maketitle

\section{Introduction}
Near-field Scanning Microwave Microscopy (NSMM) are emerging as a method to investigate microwave properties of materials at the nanoscale\cite{laiscience,rosner}. This type of scanning probe microscope can obtain information about a samples' complex microwave impedance, a quantity that not only depends on the surface properties but also gives information about  transport properties in a volume set by the extension of the microwave near-field into the sample.

It is well known that the resolution in the near-field regime can be improved well beyond the diffraction limit, and resolutions down to nanometer length scales have been demonstrated together with sub-attoFarad sensitivity, as well as loss imaging capabilities\cite{rosner, imtiaz1, imtiaz2, lai1,odagawa,gao}. 
Typically these microscopes operate at high microwave power, and at ambient conditions. However, for certain applications different regimes may be advantageous.

In fields such as circuit quantum electrodynamics, where decoherence and dissipation are major factors limiting device performance\cite{wendin}, material analysis and understanding of defects and their influence on quantum devices has become an important topic\cite{siddiqi, mcdermott}. Here an ultrasensitive microwave near-field sensor can give invaluable information on how nanoscale defects interact with macroscopic quantum devices, such as charge qubits\cite{wendin} and artificial atoms\cite{you}. NSMMs would be ideal for studying these devices since they typically operate at similar frequencies, and in combination with various other scanning probe techniques they can give substantial information on critical material properties. In addition to low temperatures, another requirement in this regime is the interaction of the probe and the sample at very low power levels, near single photon population, since quantum systems are easily saturated. 

In this paper we demonstrate a new type of cryogenic NSMM that uses a superconducting thin film resonator as the microwave sensor. Similar resonators are commonly used as building blocks in quantum circuits\cite{goppl}. Their advantage over other types of microwave resonators is that they can have very high Quality (Q) factors, up to several millions\cite{day,walraff,sage}. Naturally, a higher Q translates to increased sensitivity to variations in the microwave environment around the tip. 

Using our microscope it is possible to obtain information about both capacitance and dissipation with a relatively high sensitivity, even when the resonator is operated near the quantum limit, with a bandwidth well above that required for most scanning probe modes.
The microscope can work in parallel as a tuning-fork AFM\cite{karrai}, and we have developed a compact, low mass microwave resonator that is mounted on one of the prongs of the tuning-fork without adversely degrading the mechanical quality factor. We couple to this resonator inductively by placing a coplanar transmission line (CPW) nearby, mechanically decoupled from the tuning-fork. 

To achieve very high sensitivity we employ a measurement technique called Pound-Drever-Hall (PDH) locking, commonly used for laser stabilization\cite{black} and recently adopted to low power microwave measurements\cite{tobias}.
This technique allows to accurately track the resonance frequency of the cavity with a very high bandwidth. To demonstrate this we have used the microscope in pure NSMM-mode where tip-sample distance control uses the microwave resonance frequency as feedback signal. Scans in this mode can be acquired in the same amount of time as regular AFM scans, and on a metallic test sample we essentially see the same topography as in AFM mode. This is an indication that the microwave field is truly localized to the tip and the stray capacitance is very low. 

Furthermore, we show that we can reduce the probing microwave amplitude down to the order of $100$ $\rm{\mu}V$ (almost four orders of magnitude lower than conventional NSMMs and scanning capacitance microscopes) without considerably degrading performance and contrast. Together with non-contact AFM this becomes a very non invasive measurement. At higher powers the sensitivity is limited by mechanical noise to 64 zF/$\sqrt{\rm{Hz}}$ around typical pixel acquisition frequencies.
\section{System design}
\subsection{Microwave resonator design \& fabrication}
A superconducting quarter-wave coplanar waveguide resonator (CPWR) operating in the range 4-8 GHz has typically a length of 4-6 mm. Such a design would obviously be too bulky to integrate with an AFM force sensor. 

\begin{figure}[b!]
\begin{centering}
\includegraphics[height=8.1cm]{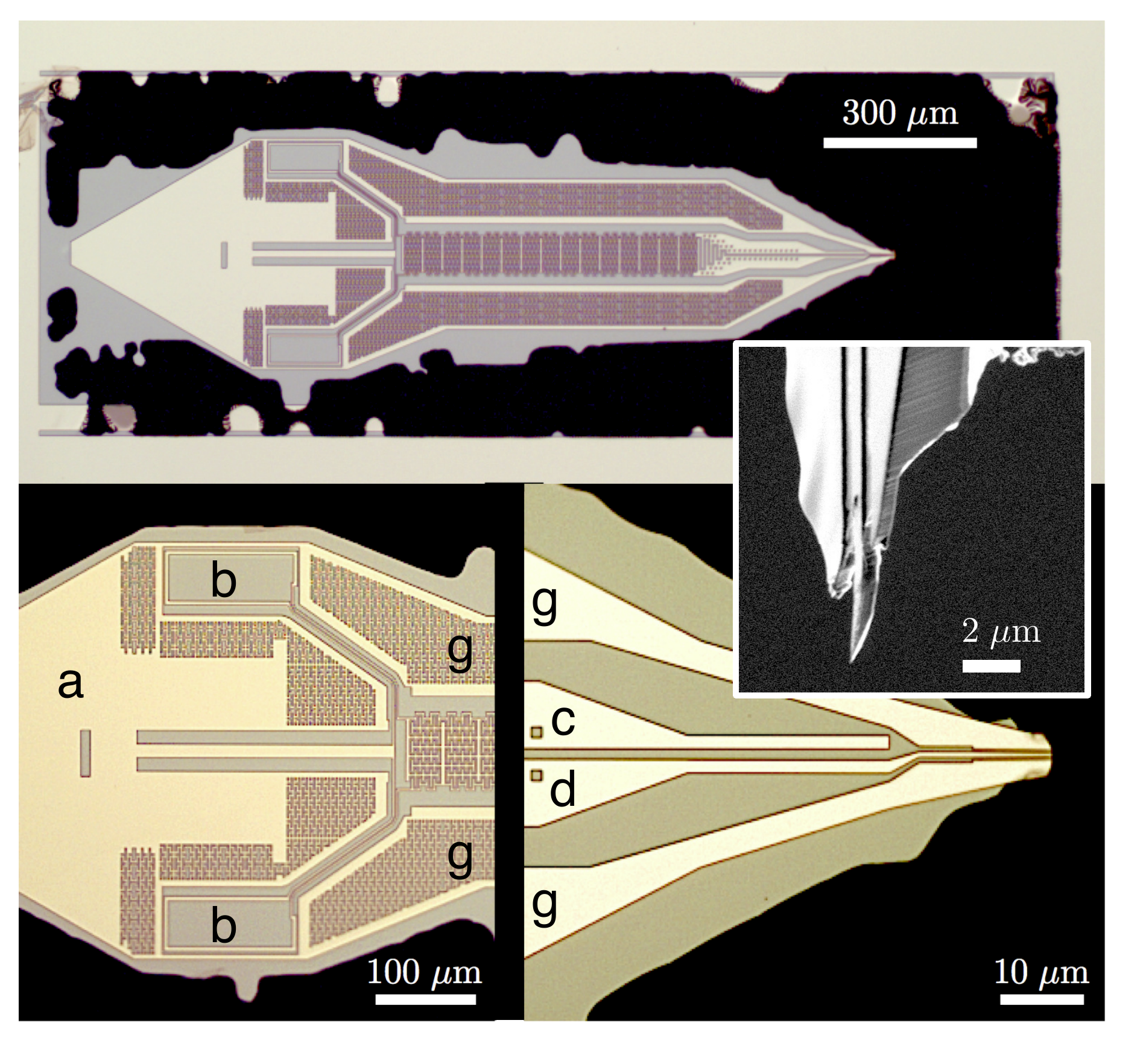}
\end{centering}
\caption{Optical image of a micromachined niobium resonator. Top: A probe as fabricated in its silicon frame. The left part consists of the inductive coupling loops while the right part shows the voltage maxima of the resonator that is terminated by the AFM tip. Bottom: Close ups of these two areas. Inset: Scanning electron micrograph taken after shaping the tip using focused ion beam. All metalized areas (a-g) are connected with 'a' being the common node.}
\label{fig:probe}
\end{figure}

For this reason we have developed a different type of resonator, shown in Fig. \ref{fig:probe}. The main idea behind this design is to reduce the size while keeping the same resonance frequency and all other properties of a distributed resonator. In fact, this resonator is conceptually similar to the mechanical tuning-fork. It consists of two 'prongs' that are loaded with a third order fractal capacitive network. The main purpose of this network is to reduce the microwave propagation velocity which leads to a reduction in length. The resonant structure supports a $\lambda/2$ standing wave, with both voltage anti-nodes close to the open end of the resonator ('c' and 'd' in Fig. \ref{fig:probe}). For a more in-depth analysis of the electrodynamics of this structure, see Ref. \onlinecite{fractalpaper}. 

We are able to considerably reduce the size of the surrounding ground planes ('g' and 'a') by reducing the capacitive coupling to the resonator. In fact, the prong-to-ground capacitance is around 1000 times smaller than the prong-to-prong capacitance, resulting in localization of the microwave currents to the resonator itself, whereas in a conventional CPWR the same ratio is close to one and the currents are close to equally distributed between the resonator and the ground. In our case the reduced currents in the ground planes mean that we can safely reduce their size without introducing spurious resonance modes and additional dissipation. However, the ground planes ('g') still serve an important function for a near-field probe: they significantly reduce stray capacitances and efficiently screen the tip. 

The resonator in Fig. \ref{fig:probe} has a few other design features that should be noted. One of the two loops close to the edge of the resonator  ('b' in Fig. \ref{fig:probe}) is used to inductively couple the resonator to an external transmission line, see Fig. \ref{fig:mainframe}. The coupling quality factor ($Q_c$) depends on the mutual inductance M between the loop and the transmission line, their respective impedances $(Z_r,Z_0)$ and the resonance frequency $\omega_0$: $Q_c = 2Z_0Z_r/\omega_0^2M^2$. 
By changing the distance between the resonator coupling loop and a nearby transmission line we can easily tune the coupling.

The second loop we keep simply to maintain the symmetry, so that we know exactly where the voltage node is located on the resonant structure\cite{note1}. At the voltage node we may connect the resonator to ground. Typically we glue a thin wire to the large metallized area shown in the left side of Fig. \ref{fig:probe} ('a') to provide a common ground for the microwave circuit that also can be DC biased relative to the sample. If the symmetry is broken there will be microwave currents induced in this resistive link to ground and the quality factor of the resonator will drop.

The resonators are fabricated on intrinsic silicon. Before sputtering 200 nm niobium the wafers are cleaned and the native oxide is removed in HF. The Nb is patterned using electron beam lithography and etching in a CF4:O2 plasma. Then the wafers are thinned down from the backside to a thickness of about 50 $\rm{\mu}$m using a standard Bosch process.  Using photolithography from the backside with infrared alignment the 'cut-out' mask is defined and the same silicon etching is repeated going through the wafer in places defined by the mask. The process provides some selectivity to niobium, and if done with enough precision it is possible to stop the silicon etch at the right moment, producing free hanging Nb that will be used as the AFM and microwave tip. 
The resonators are then mounted on one of the prongs of a quartz tuning-fork with stycast and a bonding wire is glued, using conductive cryogenic epoxy, to the large contact pad near the resonator to provide DC biasing.
To improve the sharpness of the tip beyond the film thickness we use focused ion beam to selectively mill out a sharp tip, see inset in Fig. \ref{fig:probe}. 

\subsection{Scanner design}

\begin{figure}[b!]
\begin{centering}
\includegraphics[height=11.4cm]{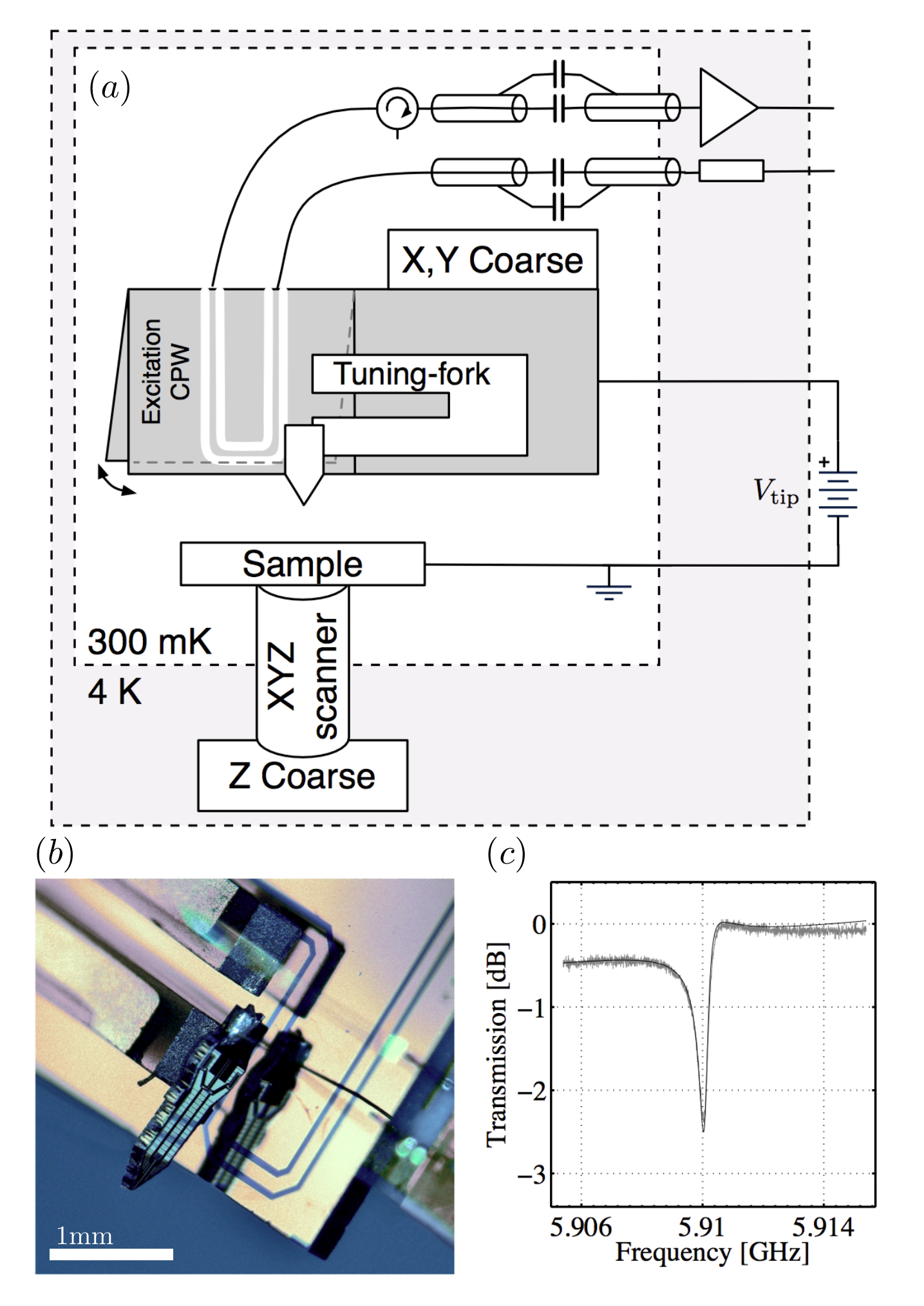}
\end{centering}
\caption{a) Schematic of the scanner setup. b) Photo of the microwave resonator mounted on the tuning-fork. Details are described in the text. c) Measured transmission ($S_{21}$) at 0.3 K of the resonator in the panel above. Fit (solid line) shows a loaded quality factor of 14600.}
\label{fig:mainframe}
\end{figure}

The microscope is assembled inside a cryostat that consists of both a nitrogen and a liquid helium vessel. The whole cryostat is suspended using pressurised air and hanging from ropes to reduce vibrations. On the 4 K plate we have mounted a compact single shot He$_3$ cooler. The scanning unit can be easily taken out, and is mounted on the 4 K plate. This unit is the main support for the various components of the scanning setup illustrated in Fig. \ref{fig:mainframe}a. The bottom part consists of an inertial slider for coarse approach (Attocube ANZ100) and two home built piezo tube scanners, one for Z and one for XY motion. On top of the Z-scanner the sample is mounted. The top part consists of two inertial sliders for XY coarse positioning and the scanner head itself supporting the tuning-fork and the microwave in/outputs. All the sliders are thermally connected to the 4 K plate, while the scanner head and the sample holder are thermally coupled to 300 mK. These two parts also have their own integrated thermometer and heater. 
Part of the assembly with the microwave resonator, tuning-fork and the microwave excitation line  is shown in Fig. \ref{fig:mainframe}b. A differential screw is used to tune the distance between the resonator and the excitation line, so that we easily can control the coupling of the microwave resonator.

The tuning-forks we use are completely removed from their original casings and glued (Epotek H77) to an alumina plate. We use wire bonding to contact the electrodes on the tuning-fork. 

Both the sample and the scanner head are electrically decoupled from the cryostat ground, as shown in Fig \ref{fig:mainframe}a. Any electrical contact is supplied through a thermalising braid attached to the 300 mK plate. In fact, the whole microwave grounding is decoupled at the 300 mK stage by DC blocks so that we can supply a DC voltage. This DC voltage will then act as a ground for the microwave sensor, but it provides an effective way of biasing the tip relative to the sample.

\subsection{Pound-Drever-Hall microwave readout}
\begin{figure}[ht!]
\begin{centering}
\includegraphics[height=15cm]{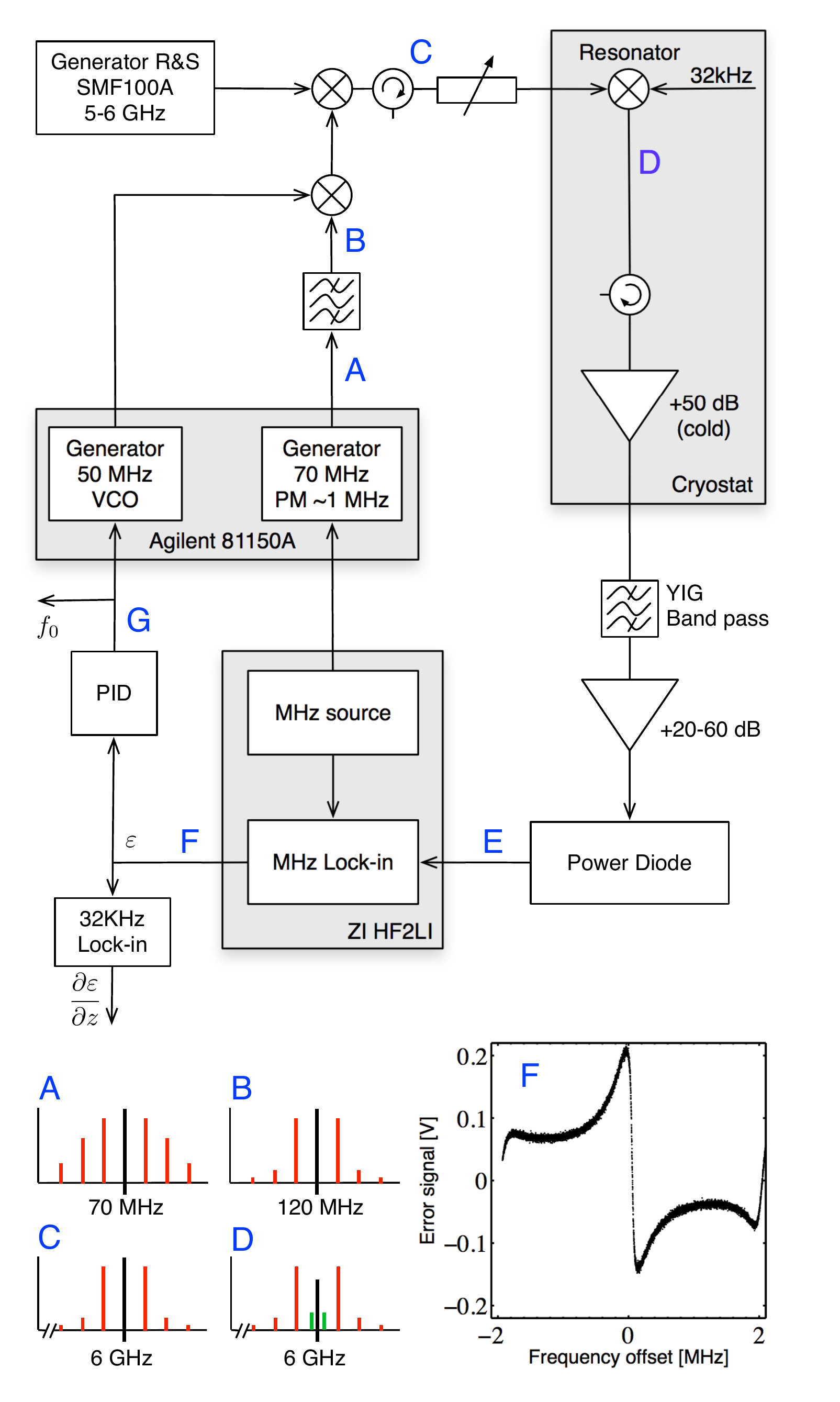}
\end{centering}
\caption{Pound-Drever-Hall microwave readout schematic. The details are discussed in the text. The bottom plots A-D sketch the microwave signal in the frequency domain at the corresponding points in the schematic. Black peak indicates the carrier frequency, red peaks the phase modulated side-band frequencies and green the additional amplitude modulation introduced by the mechanical oscillations of the AFM tuning-fork. The diode mixes the high frequency signal down to DC and the various low frequency components in the spectrum. The bottom right plot shows a typical 'error' signal measured at point F.}
\label{fig:pound}
\end{figure}
In order to increase our sensitivity and bandwidth of the microwave readout we use a technique called Pound-Drever-Hall (PDH) locking.
The essence of this method is the careful design of a probing signal spectrum that results in a signal with a very steep slope around the resonance frequency of the resonator that is probed (see Fig. \ref{fig:pound}). Using this 'error signal' as feedback to a voltage controled oscillator (VCO) it is possible to track the microwave resonance frequency with high accuracy. Rather than being limited by a maximum phase shift near resonance the 'slope' in the PDH scheme is also depending on the gain of the loop, which may be very large. This gives a very good accuracy while maintaining a large bandwidth\cite{note3}. For an in-depth analysis of this measurement technique, see for example Ref. \onlinecite{black}, and applied to superconducting thin film resonators Ref. \onlinecite{tobias}. 

Our measurement setup is illustrated in Fig. \ref{fig:pound}. We start by preparing a low frequency (70 MHz) phase modulated signal (point A in Fig. \ref{fig:pound}). The modulation frequency $\Omega$ is determined by the resonator that we want to measure. We choose the frequency such that the first side-bands are not injecting additional power into the resonator. Typically we choose a frequency of $\Omega=$1-2 MHz for a resonator with Q in the order of 20000. The optimal gain for the Pound loop is for a phase modulation depth $\beta = 1.08$\cite{black}. This results in the side-bands having a power of about -3 dB relative to the carrier, i.e. the total power in the first side-bands equals the power in the carrier. This is relatively deep phase modulation, and to reduce offsets caused by higher order side-bands and asymmetric microwave resonances we filter the signal (B) in a very narrow filter with linear phase response. The next step is to mix this signal with another signal at typically 50 MHz. This signal comes from a VCO that we later use to tune the probing frequency. Mixing is done using an ADL5391 circuit. 

Finally, (point C in Fig. \ref{fig:pound}) we do upconversion with a high frequency signal to bring our spectrum up to the frequency of the resonator, $f_0$, such that $f_{rf}+50$ MHz $+70$ MHz $=f_0$. We prepare the spectrum at low frequency for the simple reason that we can achieve much better phase stability at MHz frequencies. The power of the signal is adjusted before it is delivered into the cryostat and the resonator (D). Due to the phase shift around resonance, any deviation from the center frequency will cause the phase of the sidebands to interfere, resulting in PM-to-AM conversion\cite{black}. After sufficient amplification and filtering we send the signal to a power diode detector. 
The diode will extract the generated AM component (E).
Through a phase sensitive lock-in measurement of the $\Omega$-component we recover the signal shown at position F in Fig. \ref{fig:pound} as we sweep through the resonance frequency of the resonator. This error signal can be linearized around the resonance frequency
\begin{equation}
\varepsilon = G_{\rm{tot}}P_0J_0(\beta)J_1(\beta) Q_{\rm{tot}}(1-S_{21,\rm{min}})\frac{f-f_0}{f_0}.
\end{equation}
Here $G_{\rm{tot}}$ is the total gain of the loop, $P_0$ the total power in the signal spectrum, $J_n(\beta)$ are Bessel functions of n-th order,  $Q_{\rm{tot}}$ is the loaded quality factor and $S_{21,\rm{min}}$ is the transmission minimum (related to the coupling) of the resonator.
If we are exactly on resonance with the probing signal there is no PM-to-AM conversion and the output from the lock-in is zero. A PID controller (SRS SIM960) is then used to keep the 'error' at zero by producing a voltage that changes the frequency of the VCO. This voltage directly gives us the center frequency of the resonator (G).  In our case the bandwidth of the loop is limited by the PID which as a 3 dB roll-off at 5 kHz when tuned to its maximum bandwidth.

We also have the capability to measure the dissipation in the resonator. For this we use two different methods, depending on what we are mainly looking for in the sample. The simplest way to get an indication of the dissipation is to measure the DC component that directly comes out of the diode. This voltage has the following functional dependence on $S_{21}^{\rm{min}}$ that can be recalculated into the Q-factor.
\begin{equation}
V_{DC} = \sqrt{P_0Z_0}J_0(\beta)S_{21,\rm{min}} + C,
\end{equation}
where C is a constant coming from broad-band noise in the diode.

Another much more sensitive method is to weakly frequency modulate the carrier at a frequency above the bandwidth of the PDH loop. The error signal will have a component at this frequency which can be demodulated. I.e. we measure the slope of the error signal: $\partial\varepsilon/\partial f \propto Q_{\rm{tot}}(1-S_{21,\rm{min}})$.

\begin{figure}[b!]
\begin{centering}
\includegraphics[height=3cm]{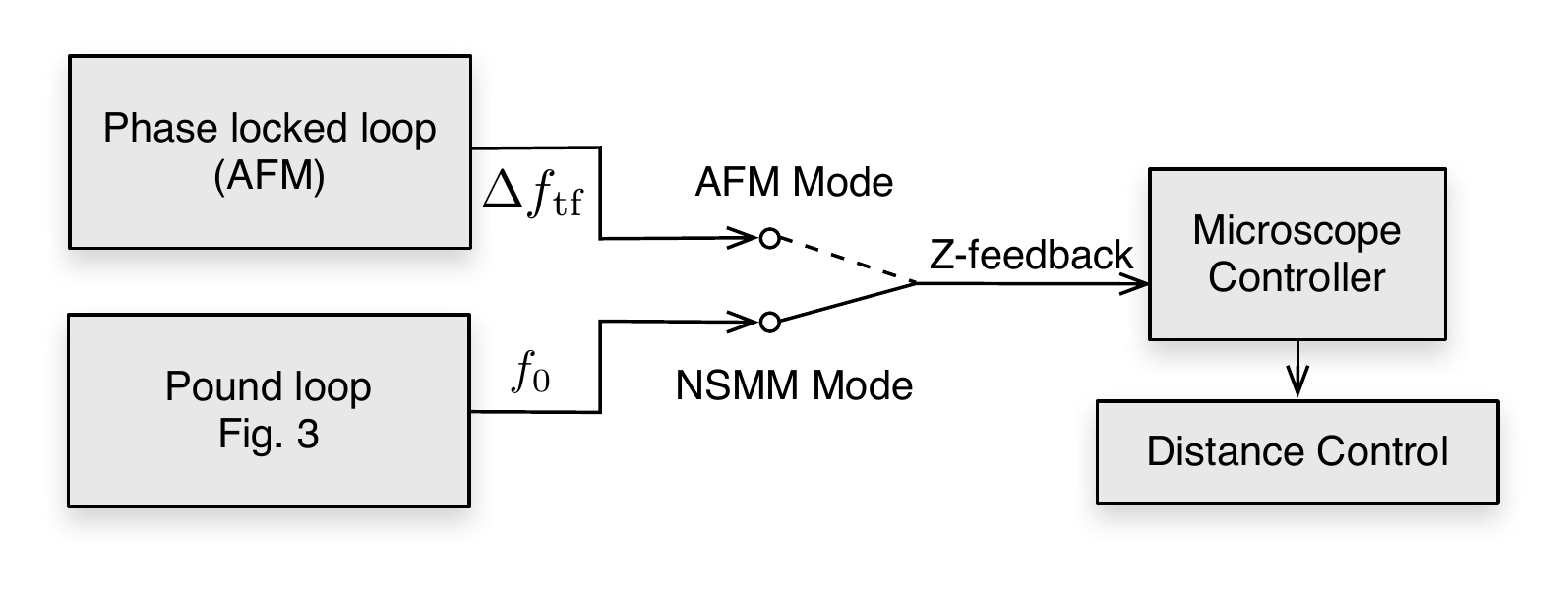}
\end{centering}
\caption{Illustration of the two different scanning modes used.}
\label{fig:feedback}
\end{figure}

The PDH scheme allows us to accurately measure the resonance frequency. We may choose to either read this out passively while scanning in normal AFM mode, or we can use the microwave frequency as the feedback signal to the microscope controller, as illustrated in Fig. \ref{fig:feedback}. 

When using the microscope in AFM mode we will also introduce an extra component in the frequency spectrum after the diode. Since the tuning fork oscillations modulate the center frequency of the resonator, 32 kHz will appear as an amplitude modulation in the final spectrum. This signal can of course be demodulated and it will give information about $df_0/dz$.

\section{Experimental results and discussion}
In Fig. \ref{fig:mainframe}c we show a typical microwave transmission of one of our probes. Typically we measure probes with internal quality factors in the range 20-30 thousands in the full assembly with DC bias connected to the resonator. At low temperatures the mechanical quality factors of the tuning forks stay above 5000 despite the additional mass added to one of the prongs.

The measured Q-factors in our scanning assembly all saturate at $Q_i\approx30000$, while we have measured $Q_i>100000$ on the same probes in another cryostat we use for pre-screening. These values are typical for Si/Nb devices\cite{barends}, to achieve even higher Q other materials have to be used. However, the fact that we loose a factor of 3 when we use the resonator in the complete scanning setup we attribute to thermal (4K) photons that generate quasiparticles in the resonator. Pumping the helium bath down to 2 K improves the quality factors by a factor $\sim$2. Introducing proper shielding at 1K around the scanner should solve this issue.

\begin{figure}[b!]
\begin{centering}
\includegraphics[height=6.2cm]{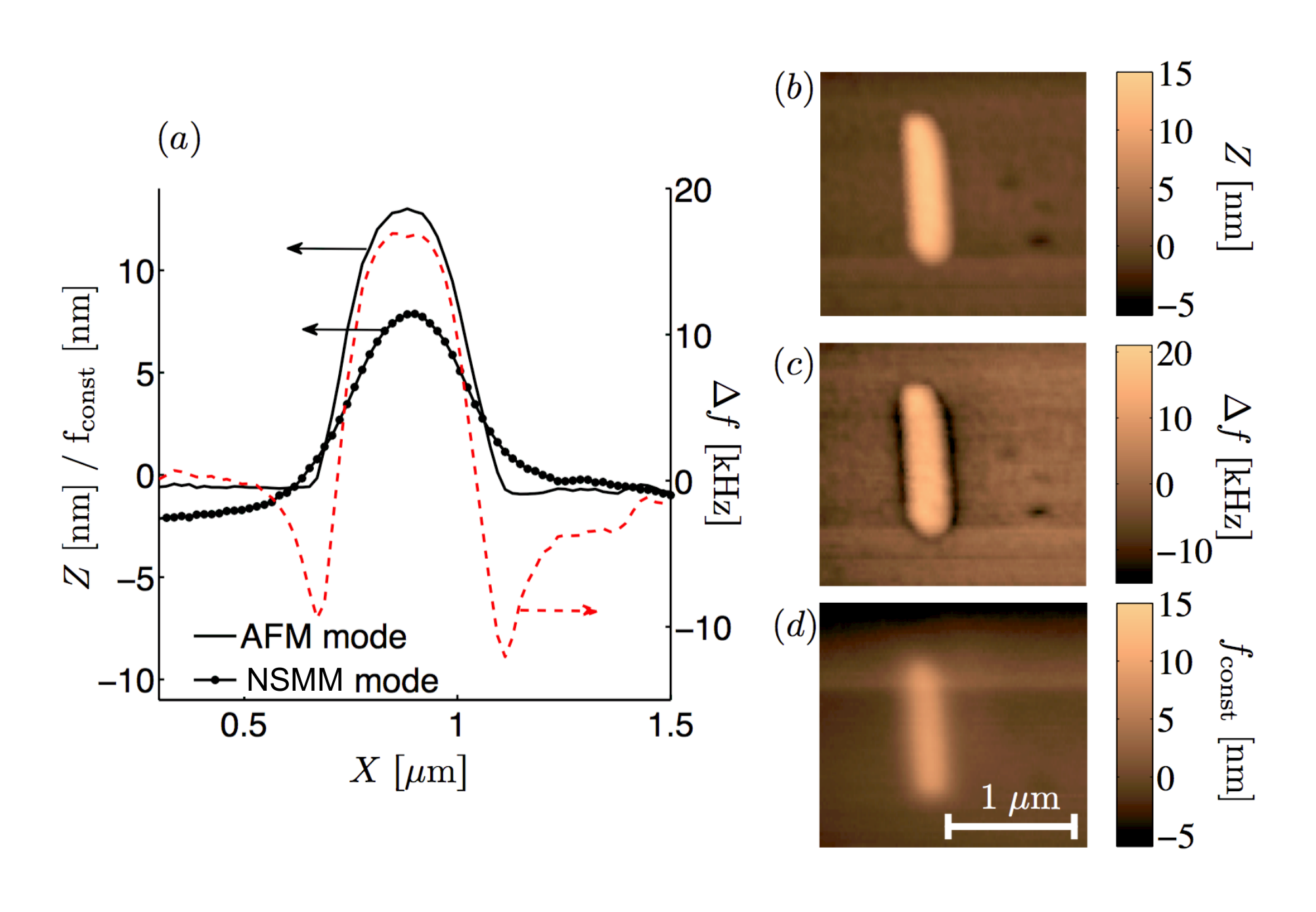}
\end{centering}
\caption{Scans over a superconducting surface with embedded topographic structures. Images obtained at 0.5 K with a scan speed of 0.8 $\mu$m/s and a microwave excitation of -80 dBm. The only post scan processing applied is a planar fit. a) Overlaid line traces from scans b-d. b) Topography in normal AFM mode. c) Microwave resonator frequency shift obtained while scanning in AFM-mode. Acquired simultaneously with b).  d) Topography obtained while using the microwave frequency as feedback signal, i.e. a surface of constant microwave resonance frequency, or constant capacitance. All other scan conditions the same as in b).}
\label{fig:fdot}
\end{figure}
\subsection{Sensitivity \& Resolution}
\begin{figure}[b!]
\begin{centering}
\includegraphics[height=5.8cm]{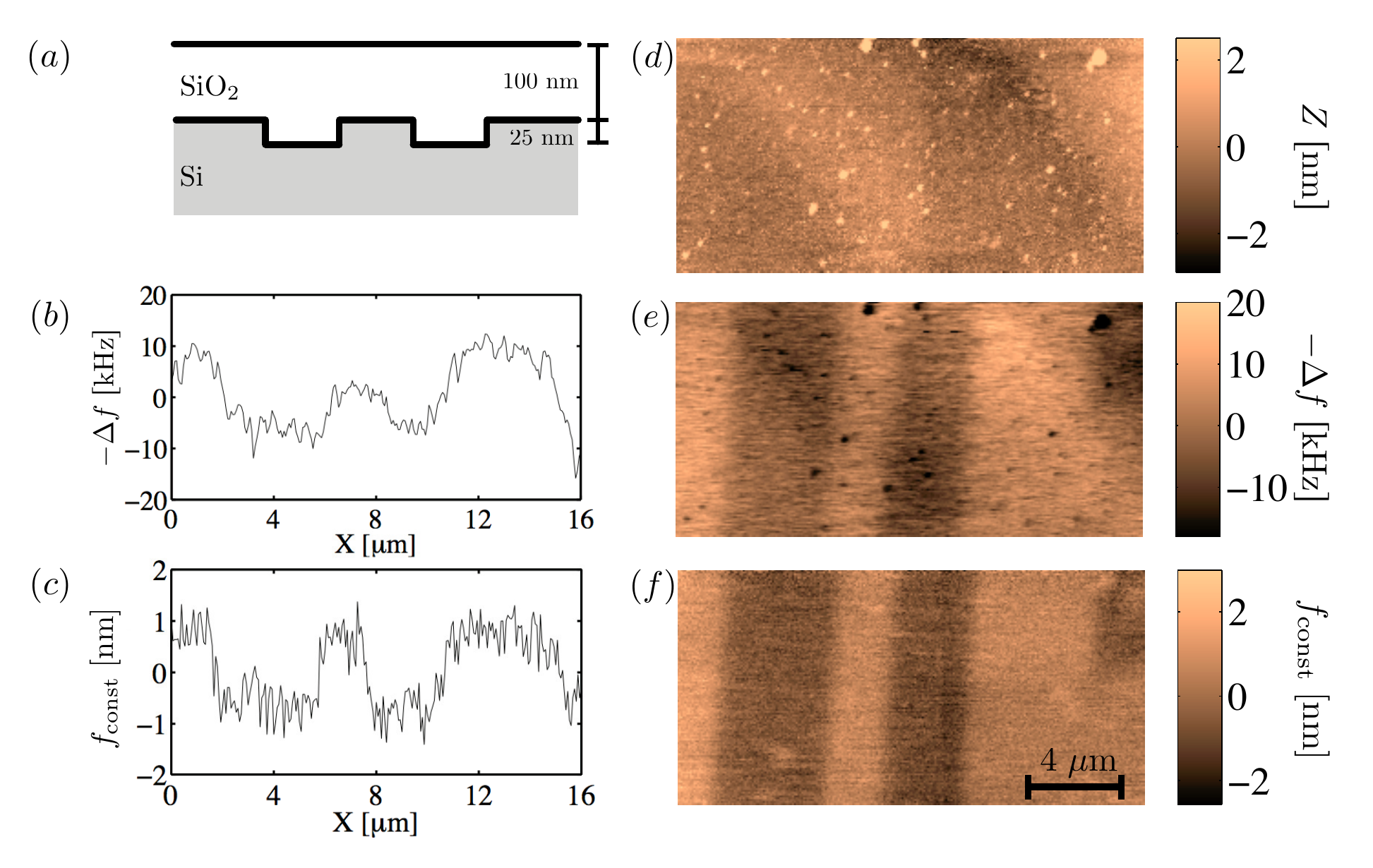}
\end{centering}
\caption{Scan over a flat dielectric sample with microfabricated variations in dielectric thickness. a) Sample cross-section. b) and c) are typical single line traces extracted from e) and f) respectively. d) AFM image of the sample surface. e) Simultaneously measured shift in resonance frequency of the microwave resonator. f) Topography acquired in NSMM mode, i.e. a surface of constant capacitance. e) is raw data, while in d) and f) a planar fit has been applied. Scans taken at 0.3 K with a microwave excitation power of -70 dBm.}
\label{fig:sitrench}
\end{figure}
One drawback for all near-field microscopes is that topography and microwave signals are heavily convoluted. A small change in topography will lift the tip and the measured capacitance and dissipation will also change accordingly. Therefore, scanning in AFM (or STM) mode over samples containing topography cannot be used to determine the true microwave resolution of a near-field microscope. Here we instead use a different approach to determine the length scale at which the microwaves are localized at the tip. By using the measured microwave resonance frequency as the active feedback signal for the distance control it is possible to determine the near-field localization relative to the AFM tip size. 

Fig. \ref{fig:fdot} shows a scan of an artificial topographic protrusion made in a superconducting film when using either AFM feedback or NSMM feedback. On the AFM scan we observe a sharp step, while on the NSMM-scan we see a slightly more smeared out topography. The microwave frequency shift obtained simultaneously with the AFM image clearly shows the effects of topography. The measured height of this protrusion is about 40\% lower in the NSMM mode, which means that nearly all of the probing field is localized to the size of the feature, i.e. $\sim 250 nm$, roughly the same size of the mechanical tip. Thus the electromagnetic field is well screened by the nearby ($\sim 3\mu$m, see inset in Fig. \ref{fig:probe}) grounding electrodes on the probe.

As another example we prepared a sample with flat topography but with dielectric contrast. The sample was fabricated by etching a pattern, 25 nm deep, with near vertical side-walls into a silicon substrate, sputtering thick silicon oxide on top and lapping the oxide thickness down to 100 nm. A sketch of the sample and the scan results are shown in Fig. \ref{fig:sitrench}. The NSMM resolution is shown here to be about 400 nm. This is what can be expected from a masking oxide thickness of 100 nm and a tip radius of around 200 nm (measured in SEM after taking the scans in Fig. \ref{fig:sitrench}d-f).  A sharper tip would result in increased resolution on the expense of a reduced capacitive contrast. The difference in capacitance between the two different regions of various thickness can thus be estimated to around 3.6 aF (assuming 200 nm radius of the tip and 10 nm tip lift). Recalculating the microwave frequency shift obtained in AFM mode into capacitance gives 3.5 aF difference. To get an estimate of the sensitivity we can calculate the signal-to-noise ratio (SNR) in Fig. \ref{fig:sitrench}c, and using the pixel sampling frequency as our bandwidth we arrive at a total sensitivity for this particular case of $66$ zF$/\sqrt{Hz}$.

A more rigorous estimation of the sensitivity in NSMM mode can be done by considering the residual frequency noise spectral density ($S_f$) of the PDH loop. In addition we have to consider the mechanical noise spectral density ($S_z$) since the mechanical instabilities are (partly) separated from the frequency measurement due to the additional feedback. Together they will give the total frequency noise
\begin{equation}
	\Delta \omega = 2\pi\left(S_f + S_z\frac{\partial f}{\partial z}\right).
\end{equation}
This can then be converted to effective capacitance noise, $\Delta C$, through
\begin{equation}
	\omega_0+\Delta\omega = \frac{1}{\sqrt{L_0(C_0+\Delta C)}},
\end{equation}
where $L_0$ and $C_0$ are calculated from the resonance frequency and resonator impedance.
 
\begin{figure}[t!]
\begin{centering}
\includegraphics[height=7.6cm]{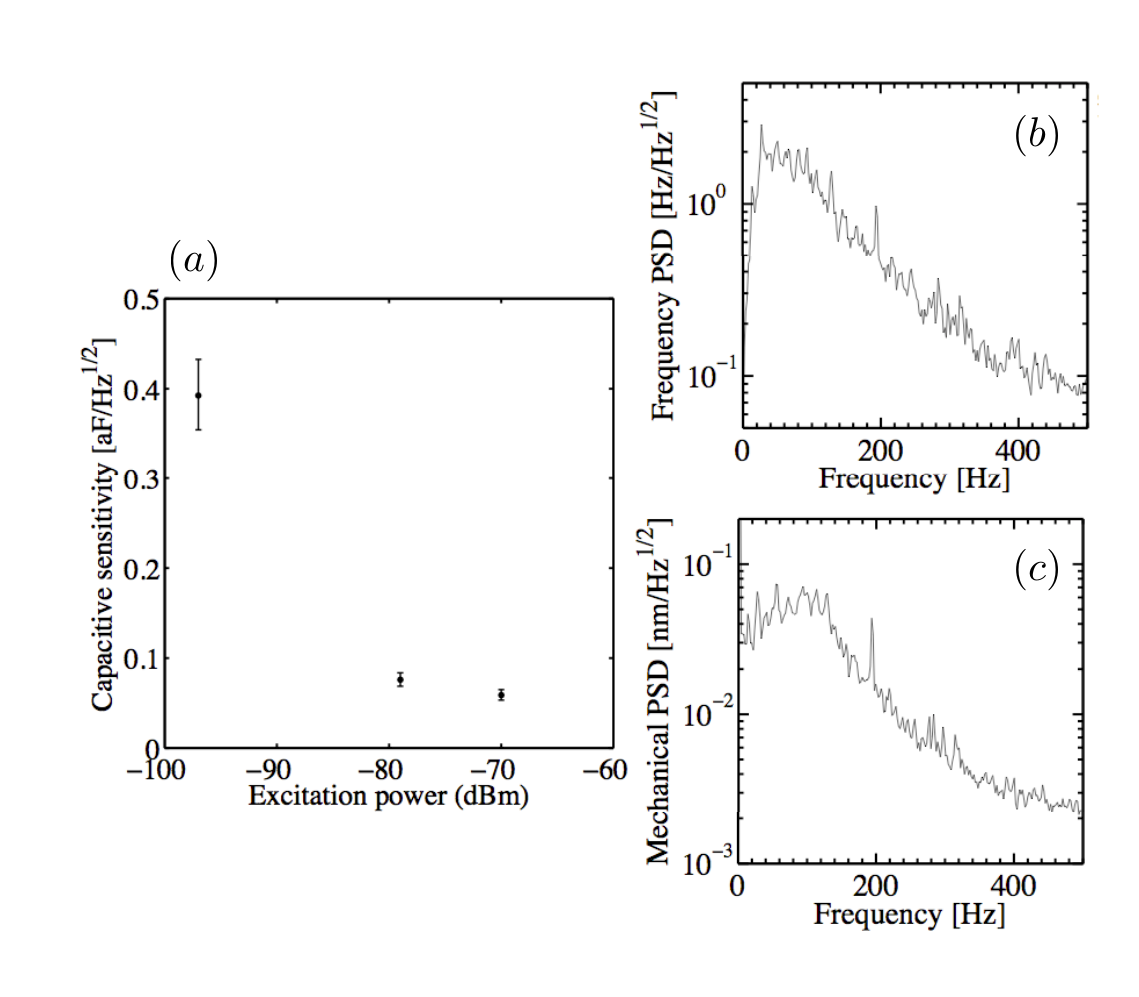}
\end{centering}
\caption{a) Capacitive sensitivity in NSMM mode for different excitation powers of the microwave resonator. -100 dBm corresponds to an energy equivalent to approximately 1000 photons for this particular resonator with $Q_i\approx16000$. Error bars are calculated asuming a fixed error in setpoint and $\partial f/\partial z$ of 5\%, plus uncertanity in mechanical z-calibration of the microscope and various other constant errors in measurements of noise spectras. Measured on top of a metallic surface. b) Microwave resonance frequency and c) vertical displacement noise power spectral density (PSD) used to calculate the sensitivity for the high power point in a).  The PID in the PDH loop is tuned to a 150 Hz bandwidth. Peak values around 50 Hz yield a capacitive sensitivity of $64 \cdot 10^{-21}$ F$/\sqrt{\rm{Hz}}$.}
\label{fig:sens}
\end{figure}

The measurement is performed by engaging the tip without scanning and measuring $S_f$ and $S_z$, a typical example of these two quantities is shown in Fig. \ref{fig:sens}b and Fig. \ref{fig:sens}c. In addition we need to know the change in frequency with respect to z-coordinate for the specific frequency set-point we choose. For this we take an approach curve, recording the frequency shift of the microwave resonator at the same time, to get a value for $\partial f/\partial z$.

Typically in this mode we achieve a microwave noise floor of 1 Hz$/\sqrt{\rm{Hz}}$ and a mechanical noise floor around 50 pm$/\sqrt{\rm{Hz}}$. At typical non contact conditions over a metallic surface (2-3 nm away) the measured frequency change is around 4 kHz/nm for the same AFM tip used earlier, of radius around 200 nm. The mechanical contribution to frequency noise is thus around 200 Hz$/\sqrt{\rm{Hz}}$. This amounts to a total capacitive sensitivity of $6.4\cdot 10^{-20}$ F$/\sqrt{\rm{Hz}}$. This figure is obtained in the high power regime (with about -70 dBm excitation on the resonator), and it is the peak noise measured at frequencies close to a typical pixel acquisition frequency while scanning. If we reduce the power the sensitivity will drop, as shown in Fig. \ref{fig:sens}a. At -100 dBm excitation (corresponding to $\sim 1000$ photons in the cavity for Q=16000\cite{note2}) the sensitivity is around $3.8\cdot 10^{-19}$ F$/\sqrt{\rm{Hz}}$. It should be noted that the noise is predominantly caused by mechanical vibrations, in a more stable scanning system the sensitivity could be improved with more than an order of magnitude. At lower powers we operate close to the noise floor of our cold amplifier ($T_N\approx 75$ K) which is the main reason for the increased noise.
In NSMM mode, we believe that the stability could be improved further by using various modulation techniques.

The residual frequency noise we measure ($\sim1$ Hz$/\sqrt{\rm{Hz}}$) is below that of a typical resonator, which often exhibits stabilities on the order of a few tens of Hz$/\sqrt{\rm{Hz}}$\cite{tobias}. Due to the additional feedback these low frequency fluctuations are translated into mechanical noise. On top of this the z-noise is the sum of several other components, including mechanical noise, thermal drift of the microwave cavity, and electronic noise of the measurement loop.
The fact that the frequency stability of the microwave resonator is less than 0.1 ppb, shows that we can be sensitive to variations in capacitance down to 0.2 zF/$\sqrt{\rm{Hz}}$, assuming no mechanical noise.

\section{Conclusion}
We have demonstrated that near-field microwave microscopy can be made truly non-invasive, both mechanically by using non-contact AFM, and electrically by operating the microwave sensor at extremely low powers. In this regime it is still possible to achieve nanometer resolution and sub-atto Farad capacitive sensitivity with a bandwidth high enough to make the microwave readout compatible with typical AFM operation. When topography is present our microscope is able to measure it in both AFM mode and NSMM mode with almost equal precision, an indication that the probing microwave field is only localized to a small volume at the apex of the tip. For typical non-contact AFM conditions we achieve a capacitive sensitivity of  0.38 aF/$\sqrt{\rm{Hz}}$ when the microwave resonator is populated with less than 1000 photons. This corresponds to a probing amplitude of less than 250 $\mu$V, four orders of magnitude lower than conventional NSMM and scanning capacitance microscopes that achieve similar sensitivity and resolution. 
For higher probing powers the sensitivity is limited to 64 zF/$\sqrt{\rm{Hz}}$ due to mechanical noise. 

This is a significant step towards future measurements of quantum systems using near-field scanning microwave probing. The performance could be further improved mainly by reducing vibrational noise,  but also by, for example, increasing the resonator quality factor and by using a microwave amplifier with lower noise temperature. \\

We aknowledge EU FP7 programme under the grant agreements  'ELFOS', the Marie Curie Initial Training Action (ITN) Q-NET 264034, the Swedish Research Council VR and the Linnaeus centre for quantum engineering for financial support.

\bibliographystyle{plainnat}

\begin{thebibliography}{26}
\bibitem{note1} Note: Since we use inductive coupling the shift in electrical length on the coupled prong becomes negligibly small.
\bibitem{note3}
Note: Another major advantage of the PDH technique, especially in cryogenic environments, is that both the signal and 'reference' (sidebands in this case) pass through the same cables. The setup is thus much less sensitivie to phase fluctuations caused by thermal drifts, and several other sources of noise are also eliminated.
\bibitem{note2} Note: The number of photons in the cavity is easily evaluated by considering the average energy inside the cavity at any given moment, $E_{res} = 2Z_0Q^2P_{exc}/\pi Z_rQ_c\omega$ and comparing this to the energy of a single photon $\hbar\omega$. Furthermore, the voltage resulting from having one photon in the cavity can easily be estimated through $\hbar\omega = CV^2/2$ where C is the capacitance of the cavity.
\end{thebibliography}

\end{document}